\documentclass[preprint]{ptptex}
\usepackage{wrapft}
\usepackage{graphicx}

\newcommand{\lsim}{{\stackrel{<}{\sim}}}
\newcommand{\kbar}{k \kern -0.5em\raise 0.6ex \hbox{--}}

\def\<{\langle}
\def\>{\rangle}
\def\bsub{\begin{subequations}}
\def\esub{\end{subequations}}
\def\beqn{\begin{eqnarray}}
\def\eeqn{\end{eqnarray}}
\def\beq{\begin{equation}}
\def\eeq{\end{equation}}
\def\b{\begin{equation}}

\title{
A Refined Numerical Result on the First Excitation Energy in the 
Two-Level Pairing Model
}

\author{
Yasuhiko {\sc Tsue},$^{1}$ 
Constan\c{c}a {\sc Provid\^encia},$^{2}$\\
Jo\~ao da {\sc Provid\^encia}$^{2}$
and Masatoshi {\sc Yamamura}$^{3}$
}

\inst{
$^1$Physics Division, Faculty of Science, Kochi University, Kochi 780-8520, 
Japan\\
$^{2}$Departamento de F\'\i sica, Universidade de Coimbra, P-3004-516 Coimbra, 
Portugal\\
$^{3}$Faculty of Engineering, Kansai University, Suita 564-8680, Japan
}

\recdate{
\today
}

\abst{
The first excitation energy in the two-level pairing model is investigated in 
terms of the equilibrium and the small fluctuation around it. 
The basic idea is an extension of results presented in a previous paper 
by the present authors. 
In this investigation, the result obtained in the previous paper plays a 
central role: 
At the limit of the weak and the strong interaction strength, the results are 
in good agreement with the exact one. 
The former and the latter are calculated in the framework of 
the $su(2)\otimes su(1,1)$- and the $su(2)\otimes su(2)$-coherent state, 
respectively. 
On the basis of the above conclusion, the intermediate region for the 
interaction strength is described in terms of the idea of the 
interpolation and it is shown that the agreement of the result with the exact 
one is quite good. 
}
\begin{document}
\maketitle


The description of the first excited state in the two-level pairing model 
for the closed shell system has been regarded as a basic, but a modest 
problem in many-body physics. 
For example, the recent investigations can be found in Ref. \citen{1}. 
Following this quiet stream, recently, the present authors have 
considered this problem from tow slightly different viewpoints. 
One was reported in Ref.\citen{2} and the other in Refs.\citen{3} and 
\citen{4}. 
Especially, in Ref.\citen{4} (referred to as (I)), 
the ground-state and the first excitation energy are calculated 
in the framework of three different coherent states which consisit 
of four kinds of boson operators. 
In (I), they are called (i) the $su(2)\otimes su(1,1)$-, 
(ii) the $su(1,1)\otimes su(1,1)$- 
and (iii) the $su(2)\otimes su(2)$-coherent states. 
The picture adopted in (I) is based on the equilibrium and the 
fluctuation around it and the equilibrium state 
is described by a coherent state. 
Therefore, we obtain three different equilibrium states. 
Roughly speaking, for the first excited energy, 
in the region of weak strength of the pairing interaction, 
the state (i) gives rather good result. 
But, in the region of strong strength, the state (iii) gives good result. 
Compared with the above two states, 
it seems to us that the state (ii) gives a slightly worse result. 
Of course, the above rough summary is based on the comparison with 
the exact result.

Before entering the central parts of this paper, we mention 
our basic viewpoint. 
We pay attention to the relation appearing in the 
appendix (B) of (I), i.e., (I$\cdot$B$\cdot$1). 
This relation demonstrates that, if we solve our system exactly 
in the quantum mechanical framework, the result does not depend on the choice 
of the function $g({\hat K})$. 
The part of the Hamiltonian, ${\hat H}_1$, is recast into 
\beqn\label{1}
{\hat H}_1&=&
-G\left[\sqrt{g({\hat K})}\cdot\sqrt{\hbar}{\hat c}^*\sqrt{2T+{\hat K}}
\sqrt{2M-{\hat K}}\sqrt{2L-{\hat K}}\cdot
\left(\sqrt{g({\hat K}+\hbar)}\right)^{-1} +{\rm h.c.}\right] \nonumber\\
&=&-G\left[\sqrt{\hbar}{\hat c}^*\sqrt{2T+{\hat K}}\sqrt{2M-{\hat K}}
\sqrt{2L-{\hat K}}+{\rm h.c.}\right] \ ,
\eeqn
where h.c. means the Hermitian conjugation. 
It should be noted that the relation (\ref{1}) holds for any form of 
$g({\hat K})$ if it can be defined. 
On the other hand, the $c$-number replacement gives the form 
\beqn\label{2}
{H}_1&=&
-G\left[\sqrt{g(K)}\cdot\sqrt{\hbar}{c}^*\sqrt{2T+{K}}
\sqrt{2M-{K}}\sqrt{2L-{K}}\cdot
\left(\sqrt{g(K+\hbar)}\right)^{-1} +{\rm c.c.}\right] \nonumber\\
&=&-G\left[\sqrt{\hbar}{c}^*\sqrt{2T+{K}}\sqrt{2M-{K}}
\sqrt{2L-{K}}\cdot\sqrt{g(K)\cdot g(K+\hbar)^{-1}}+{\rm c.c.}\right] \ ,
\eeqn
where c.c. means the complex conjugation. 
The relation (\ref{2}) tells us that the classical solution based on the 
relation (\ref{2}) depends on the choice of $g(K)$. 
The above means that the classical counterpart under the 
$c$-number replacement cannot be fixed uniquely, 
that is, there exist infinite possibilities for the classical solutions 
which should correspond to the unique quantum mechanical solution. 
However, we must note that we determine the equilibrium by a classical 
solution and the fluctuation is given in the lowest order for 
classical and quantum mechanical framework. 
Therefore, the result depends on the form of $g(K)$, and the principle, 
with the aid of which $g(K)$ is fixed, is required. 
In (I), we investigated three possibilities which come from three forms of the 
coherent state.

As was demonstrated in (I), the key to obtain a good result is to choose 
the function $f({\hat K})$ related to $g({\hat K})$ and 
defined through the form (I$\cdot$B$\cdot$2), i.e., $f(K)$. 
The three cases are shown in the 
relation (I$\cdot$4$\cdot$4) and in Fig.(I$\cdot$2), 
the different behaviors are explicitly presented. 
Here, ``good result" means the agreement with the exact one. 
From this figure, we learn that if we succeed in finding the 
function $f(K)$, which satisfies the following condition, the good result may 
be expected: 
For a small value of $K$, $f(K)$ shows the behavior shown in 
Fig.(I$\cdot$2a) and in the region $K\lsim \Omega/2$, 
the behavior of $f(K)$ should 
be as shown in Fig.(I$\cdot$2c). 
In this paper, we present some candidates for $f(K)$ 
which show the above behavior under the idea of an interpolation.

Instead of $f(K)$, we investigate the function $\mu(K)$ defined as 
\beqn\label{3}
& &f(K)=2G\mu(K) \ .
\eeqn
In the case of the states (i) and (iii), $\mu(K)$ can be expressed 
in the form 
\beqn
& &\mu_{\rm i}(K)=\sqrt{K(K+1)}(\Omega-K) \ , 
\label{4}\\
& &\mu_{\rm iii}(K)=K(\Omega-K) \ . 
\label{5}
\eeqn
Hereafter, we use the unit $\hbar=1$. 
The function $\mu_{\rm i}(K)$ can be expanded in the form 
\beqn\label{6}
& &\mu_{\rm i}(K)=\sqrt{K}\left(\Omega+(\Omega/2-1)K+ \cdots \right) \ .
\eeqn
On the other hand, $\mu_{\rm iii}(K)$ can be rewritten as 
\beqn\label{7}
& &\mu_{\rm iii}(K)=(\Omega/2)^2+0\times (\Omega/2-K)-
(\Omega/2-K)^2 \ . 
\eeqn
Therefore, the leading terms for $\mu_{\rm i}(K)$ near $K\sim 0$ 
and for $\mu_{\rm iii}(K)$ near $K\sim \Omega/2$ are as follows: 
\beqn
& &\mu_{\rm i}(K)=\sqrt{K}\Omega \ , 
\label{8}\\
& &\mu_{\rm iii}(K)=(\Omega/2)^2+0\times (\Omega/2-K) \ . 
\label{9}
\eeqn
Our problem is to find an explicit form for $\mu(K)$ which reduces to the 
forms (\ref{8}) and (\ref{9}) for $K\sim 0$ and $K\sim \Omega/2$, 
respectively, in the framework of the idea of an interpolation.

First, let us search $\mu(K)$ in the form 
\beqn\label{10}
& &\mu(K)=\sqrt{K(K+n)}A(B-K) \ .
\eeqn
Here, $n$, $A$ and $B$ denote parameters to be determined as functions 
of $\Omega$. 
The function $\mu(K)$ is expanded in the form 
\bsub\label{11}
\beqn
& &\mu(K)=\sqrt{K}\left[\sqrt{n}AB+A(B/2\sqrt{n}-\sqrt{n})K+ \cdots \right] \ .
\label{11a}
\eeqn
Also, in terms of $(\Omega/2-K)$, $\mu(K)$ is expressed as 
\beqn
\mu(K)&=&\sqrt{\Omega(\Omega+2n)}/2\cdot A(B-\Omega/2) \nonumber\\
& &+\sqrt{\Omega(\Omega+2n)}/2\cdot A\left[
1-(B-\Omega/2)\frac{2(\Omega+n)}{\Omega(\Omega+2n)}\right](\Omega/2-K) 
\nonumber\\
& &-\frac{A}{\sqrt{\Omega(\Omega+2n)}}\left[
(\Omega+n)+(B-\Omega/2)\frac{n^2}{\Omega(\Omega+2n)}\right](\Omega/2-K)^2
+\cdots \ . \nonumber\\
& &\label{11b}
\eeqn
\esub
Comparison of the forms (\ref{11a}) and (\ref{11b}) with the asymptotic 
forms (\ref{8}) and (\ref{9}) gives us the relation 
\bsub\label{12}
\beqn
& &\sqrt{n}AB=\Omega \ , 
\label{12a}\\
& &\sqrt{\Omega(\Omega+2n)}A(B-\Omega/2)=(\Omega/2)^2 \ , 
\label{12b}\\
& &1-(B-\Omega/2)\cdot\frac{2(\Omega+n)}{\Omega(\Omega+2n)}=0 \ . 
\label{12c}
\eeqn
\esub
The relations (\ref{12b}) and (\ref{12c}) lead to 
\beqn\label{13}
& &A=\frac{\sqrt{\Omega}(\Omega+n)}{\left(\sqrt{\Omega+2n}\right)^3} \ , 
\qquad
B=\Omega\left(\frac{\Omega+3n/2}{\Omega+n}\right) \ . 
\eeqn
Substituting the form (\ref{13}) into the relation (\ref{12a}), 
we have 
\beqn\label{14}
n=\frac{(\Omega+2n)^3}{\Omega(\Omega+3n/2)^2} \ . 
\eeqn
By solving the relation (\ref{14}), we can express $n$ in terms of $\Omega$ 
and the relation (\ref{13}) gives $A$ and $B$ expressed in terms of 
$\Omega$. 
In the case $\Omega=5$, for which we show the numerical results in figures, 
we have 
\beqn\label{15}
& &n=2.8441 \ , \qquad A=0.5020 \ , \qquad B=5.9064 \ .
\eeqn

Next, we investigate a possible improvement of the form (\ref{10}) with 
the relations (\ref{13}) and (\ref{14}). 
We set up the following form: 
\beqn\label{16}
\mu(K)=\sqrt{K(K+n)}A(B-K)\cdot \rho(K) \ .
\eeqn
If $\rho(K)=1$ in the region $0 \leq K \leq \Omega/2$, the 
improvement is not achieved. 
If we intend to keep the forms (\ref{13}) and (\ref{14}) in the 
improvement, $\rho(K)$ should satisfy 
\beqn\label{17}
& &\rho(0)=\rho(\Omega/2)=1 \ , \qquad 
\rho'(\Omega/2)=0 \ .
\eeqn
As a possible candidate, we can choose the form 
\beqn\label{18}
& &\rho(K)=\sqrt{\frac{1+C\cdot(\Omega/2)^{-4}(\Omega/2-K)^4}
{1+C\cdot(\Omega/2)^{-2}(\Omega/2-K)^2}} \ .
\eeqn
Here, $C$ denotes a new parameter to be determined as a function of 
$\Omega$. 
From the comparison with the exact result, we can see that the large value 
of $K$ $(\sim \Omega/2)$ should be improved. 
Then, we expand $\rho(K)$ in terms of $(\Omega/2-K)$: 
\beqn\label{19}
& &\rho(K)=1-(C/2)\cdot (\Omega/2)^{-2}(\Omega/2-K)^2+\cdots \ . 
\eeqn
Then, the expression (\ref{11b}) should be improved in the form 
\beqn\label{20}
& &\hbox{\rm the term improved from the third term in the expansion 
(\ref{11b})} \nonumber\\
&=& -\biggl[\frac{A}{\sqrt{\Omega(\Omega+2n)}}\left((\Omega+n)
+(B-\Omega/2)\frac{n^2}{\Omega(\Omega+2n)}\right) \nonumber\\
& &+\sqrt{\Omega(\Omega+2n)}/2\cdot A(B-\Omega/2)\cdot (C/2)(\Omega/2)^{-2}
\biggl](\Omega/2-K)^2 \ . 
\eeqn
By putting the form (\ref{20}) equal to the third terms of the 
relation (\ref{7}), we obtain 
\beqn\label{21}
& &\frac{A}{\sqrt{\Omega(\Omega+2n)}}\left[(\Omega+n)
+(B-\Omega/2)\frac{n^2}{\Omega(\Omega+2n)}\right] \nonumber\\
& &+\sqrt{\Omega(\Omega+2n)}/2\cdot A(B-\Omega/2)\cdot (C/2)(\Omega/2)^{-2}
=1 \ .
\eeqn
By substituting the result (\ref{13}) for $A$ and $B$ into the form 
(\ref{21}), $C$ is determined as 
\beqn\label{22}
& &C=\frac{2n(\Omega+5n/4)}{(\Omega+2n)^2} \ . 
\eeqn
The case $\Omega=5$ gives us 
\beqn\label{23}
& &C=0.4260 \ .
\eeqn

As is clear from the relation (I$\cdot$B$\cdot$4), it is also an important 
task for the present description to obtain the function $g(K)$ defined through 
the relation 
\beqn\label{27}
& &\mu(K)=\sqrt{K(K+1)}(\Omega-K)\sqrt{g(K)\cdot g(K+1)^{-1}} \ .
\eeqn
In the present case, we have 
\beqn\label{28}
& &g(K)\cdot g(K+1)^{-1}=
A^2\left(\frac{K+n}{K+1}\right)\left(\frac{B-K}{\Omega-K}\right)
\cdot \xi(\Omega, K) \ . 
\eeqn
For the expression (\ref{10}), we have 
\beqn\label{29}
& &\xi_1(\Omega, K)=1 \ . 
\eeqn
For the expression (\ref{16}) with the candidate (\ref{18}), 
$\xi_2(\Omega, K)$ is given as 
\beqn\label{30}
\xi_2(\Omega, K)&=&(\Omega/2)^{-2}
\left[(\Omega/2)\left\{(1+1/\sqrt{2}\sqrt[4]{C})+i\cdot 1/\sqrt{2}\sqrt[4]{C}
\right\}-K\right] \nonumber\\
& &\qquad\qquad
\times\left[(\Omega/2)\left\{(1-1/\sqrt{2}\sqrt[4]{C})
+i\cdot 1/\sqrt{2}\sqrt[4]{C}\right\}-K\right] \nonumber\\
& &\qquad\qquad
\times
\left[(\Omega/2)(1+i/\sqrt{C})-K\right]^{-1} \nonumber\\
& &\qquad\qquad
\times \left[{\rm c.c.}\right] \ .
\eeqn
The notation $[{\rm c.c}]$ denotes the complex conjugate of all the 
previous terms. 
Hereafter, we distinguish the two cases by the indeces 1 and 2. 
For finding the functions $g_1(K)$ and $g_2(K)$, the following 
formulae are useful: 
\bsub\label{32}
\beqn
&(i)&\ {\rm if}\quad h(K)\cdot h(K+1)^{-1}=\alpha\ , \qquad 
h(K)=\beta\alpha^{-K} \ , 
\label{32a}\\
&(ii)&\ {\rm if}\quad h(K)\cdot h(K+1)^{-1}=\alpha -K\ , \qquad 
h(K)=\beta\Gamma(\alpha+1-K) \ , 
\label{32b}\\
&(iii)&\ {\rm if}\quad h(K)\cdot h(K+1)^{-1}=\alpha +K\ , \qquad 
h(K)=\beta\Gamma(\alpha+K)^{-1} \ . 
\label{32c}
\eeqn
\esub
Here, $\alpha$ and $\beta$ do not depend on $K$ and $\Gamma(z)$ denotes the 
gamma-function. 
As can be seen in the relation (I$\cdot$B$\cdot$4), we need only the form 
$g'(K_0)/g(K_0)$, and then, it may be enough to put $\beta=1$ for 
the formula (\ref{32}). 
With the use of the formula (\ref{32}), we obtain the following relations 
for $g(K)$: 
\beqn
g_1(K)&=&
A^{-2K}\cdot \Gamma(1+K)\cdot \Gamma(n+K)^{-1}\cdot 
\left[\Gamma(B+1-K)\cdot \Gamma(\Omega+1-K)^{-1}\right]^2 \ , \qquad
\label{33}\\
g_2(K)&=&
g_1(K)(\Omega/2)^{2K} \nonumber\\
& &\times \Gamma((\Omega/2)\{(1+1/\sqrt{2}\sqrt[4]{C})
+i\cdot 1/\sqrt{2}\sqrt[4]{C}\}+1-K) \nonumber\\
& &\times \Gamma((\Omega/2)\{(1-1/\sqrt{2}\sqrt[4]{C})
+i\cdot 1/\sqrt{2}\sqrt[4]{C}\}+1-K) \nonumber\\
& &\times \left[\Gamma((\Omega/2)(1+i/\sqrt{C})+1-K)\right]^{-1} \nonumber\\
& &\times \left[ {\rm c.c.} \right] \ . 
\label{34}
\eeqn

%
\begin{figure}[t]
\begin{center}
\includegraphics[height=5.8cm]{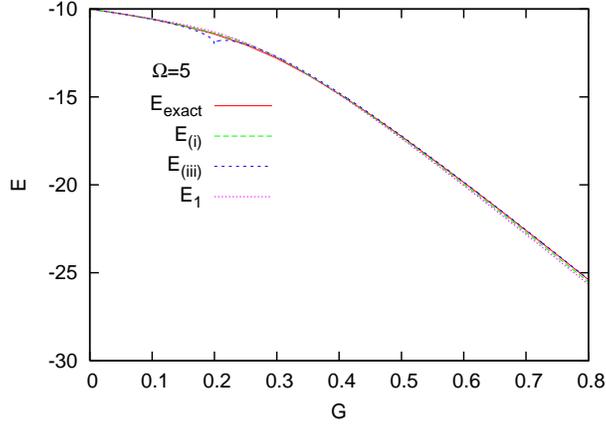}
\caption{The energies obtained by the use of the various states are 
depicted as a function of 
$G$. Here, 
$E_{\rm (i)}$ and $E_{\rm (iii)}$ denote the derived energies by using of 
$\mu_{\rm i}$ and $\mu_{\rm iii}$ in Eqs.(\ref{4}) and (\ref{5}), respectively, and $E_{1}$ denotes the energy by using of $\mu(K)$ in (\ref{10}) with 
(\ref{13})$\sim$(\ref{15}), where $\Omega=5$. 
Here, $E_{\rm exact}$ shows the exact eigenvalue of the Hamiltonian. 
}
\label{fig:1}
\end{center}
\end{figure}
%
%
%
%
\begin{figure}[t]
\begin{center}
\includegraphics[height=5.8cm]{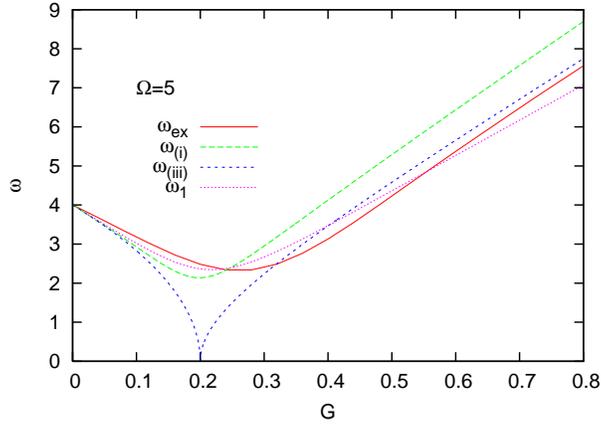}
\caption{The frequencies are depicted as a function of $G$. Here, 
$\omega_{\rm (i)}$ and $\omega_{\rm (iii)}$ denote the derived 
frequencies by using of 
$\mu_{\rm i}$ and $\mu_{\rm iii}$ in Eqs.(\ref{4}) and (\ref{5}), respectively, and $\omega_{1}$ denotes the frequency by using of $\mu(K)$ in (\ref{10}) with 
(\ref{13})$\sim$(\ref{15}), where $\Omega=5$. 
Here, $\omega_{\rm ex}$ shows the exact result. 
}
\label{fig:2}
\end{center}
\end{figure}
%

%
\begin{figure}[t]
\begin{center}
\includegraphics[height=5.8cm]{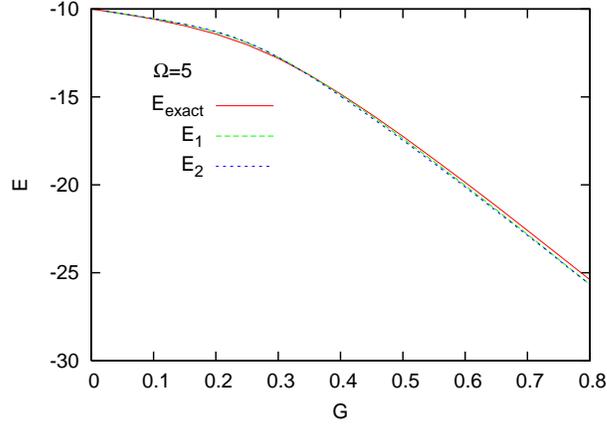}
\caption{The energies obtained by the use of the new $\mu(K)$ are 
depicted as a function of 
$G$. Here, 
$E_{1}$ and $E_{2}$ denote the derived energies by using of 
$\mu(K)$ in (\ref{10}) and (\ref{16}) with (\ref{18}), respectively, 
and $E_{\rm exact}$ shows the exact 
eigenvalue of the Hamiltonian. Here, $\Omega=5$. 
}
\label{fig:3}
\end{center}
\end{figure}
%
%
\begin{figure}[t]
\begin{center}
\includegraphics[height=5.8cm]{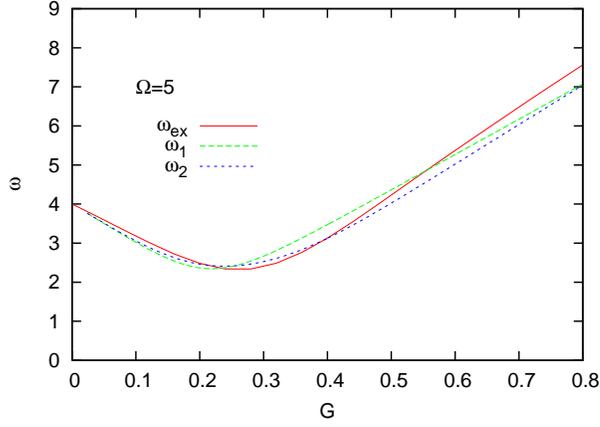}
\caption{The frequencies obtained by the use of the new $\mu(K)$ are 
depicted as a function of 
$G$. Here, 
$\omega_{1}$ and $\omega_{2}$ denote 
the derived frequencies by using of 
$\mu(K)$ in (\ref{10}) and (\ref{16}) with (\ref{18}), respectively, 
and $\omega_{\rm ex}$ shows the exact 
result. Here, $\Omega=5$. 
}
\label{fig:4}
\end{center}
\end{figure}
%

In Fig.\ref{fig:1}, the derived energies of equilibrium are compared with the 
exact ground state energy $E_{\rm exact}$. Here, $E_{\rm (i)}$ and 
$E_{\rm (iii)}$ denote the energies obtained by using $\mu_{\rm i}$ 
in Eq.(\ref{4}) and $\mu_{\rm iii}$ in Eq.(\ref{5}), respectively, 
and $E_{1}$ denotes the energy newly obtained 
by using the $\mu(K)$ in Eq.(\ref{10}). 
Also, in Fig.\ref{fig:2}, the derived first excitation energies or frequencies 
of small oscillations around the equilibrium are compared with the 
exact first excitation energy $\omega_{\rm ex}$. Here, $\omega_{\rm (i)}$ 
and 
$\omega_{\rm (iii)}$ denote the first excitation energies obtained 
by using $\mu_{\rm i}$ 
in Eq.(\ref{4}) and $\mu_{\rm iii}$ in Eq.(\ref{5}), respectively, 
and $\omega_{1}$ denotes the first excitation 
energy newly obtained 
by using the $\mu(K)$ in Eq.(\ref{10}). 
The energy of equilibrium is not so refined except for the region 
presenting a 
dip structure in $E_{\rm (iii)}$. 
However, the frequency $\omega_1$ is obviously refined. 
In the region of the weak interaction strength $G$, the behavior of 
$\omega_1$ is similar to $\omega_{\rm (i)}$ by using the $su(2)\otimes 
su(1,1)$-coherent state as is expected. 
On the other hand, 
in the region of the strong interaction strength, the behavior of 
$\omega_1$ is similar to $\omega_{\rm (iii)}$ by using the $su(2)\otimes 
su(2)$-coherent state. 
As a result, the obtained result is better in comparison with the exact 
first excitation energy.

Further, we have improved the function $\mu(K)$ by introducing the 
function $\rho(K)$ in Eq.(\ref{10}). 
Here, we attach the suffix 2 to the quantities derived by 
using the function $\rho(K)$ in Eqs.(\ref{18}). 
In Fig.3, 
the equilibrium energies are depicted in order to compare them with the exact 
ground state energy and $E_1$ corresponding to the case $\rho(K)=1$. 
The refinement by using the newly introduced function $\rho(K)$ is also 
seen in the frequency $\omega$ in Fig.\ref{fig:4} transparently. 
In the weak interaction strength region, the refinement is further obtained 
in comparison with $\omega_1$ with $\rho(K)=1$. 
In the strong intearction strength region, the obtained frequency runs 
almost parallel to the exact one as a function of $G$. 
In addition to the above-mentioned refinement, 
even in the intermediate region of the interaction strength, 
the behavior is similar to the exact result and a fairly good result 
is obtained.

In conclusion, the refined results for 
the 
first excitation energy around the equilibrium given by the coherent states 
have been obtained in the formulation of the idea in which 
the function $g(K)$ reveals 
the similar asymptotic behavior to those of the $su(2)\otimes su(1,1)$- 
and the $su(2)\otimes su(2)$-coherent states. 
The results shown in this paper teaches us that, following the 
increase of the interaction strength, the $su(2)\otimes su(1,1)$-symmetry 
changes gradually to the $su(2)\otimes su(2)$-symmetry. 
An interesting future problem is to investigate which coherent state 
can describe the above-mentioned change of the symmetry.

\section*{Acknowledgements} 
This work started when two of the authors (Y. T. and M. Y.) 
stayed at Coimbra in September 2005 and was completed when 
M. Y. stayed at Coimbra in August and September 2006. They would like to 
express their sincere thanks to Professor Jo\~ao da Provid\^encia, a 
co-author of this paper, for his invitation and warm hospitality. 
One of the authors (Y. T.) 
is partially supported by a Grant-in-Aid for Scientific Research 
(No.15740156 and No.18540278) 
from the Ministry of Education, Culture, Sports, Science and 
Technology of Japan. 




\begin{thebibliography}{99}
\bibitem{1}
M. Samabataro and N. Dinh Dang, Phys. Rev. C {\bf 59} (1999), 1422.\\
A. Rabhi, R. Bennaceur, G. Chanfray and P. Schuck, Phys. Rev. C {\bf 66} 
(2002), 064315.\\
N. Dinh Dang, Euro. Phys. J. A {\bf 16} (2003), 181.\\
A Rabhi, Eur. Phys. J. A {\bf 20} (2004), 277. 
\bibitem{2}
M. Yamamura, C. Provid\^encia, J. da Provid\^encia, F. Cordeiro and Y. Tsue, 
J. of Phys. A: Math. Gen. {\bf 39} (2006), 11193.
\bibitem{3}
C. Provid\^encia, J. da Provid\^encia, Y. Tsue and M. Yamamura, 
Prog. Theor. Phys. {\bf 115} (2006), 739; 759.
\bibitem{4}
Y. Tsue, C. Provid\^encia, J. da Provid\^encia and M. Yamamura, 
submitted to Prog. Theor. Phys. (nucl-th/0610074)
\end{thebibliography}
\end{document}